\documentclass[showpacs,amsmath,amssymb,aps,twocolumn,superscriptaddress,prx]{revtex4-2} 
\usepackage{mathtools}
\usepackage{algorithm}
\usepackage{algpseudocode}
\usepackage[pdftex]{graphicx} \graphicspath{{}}
\usepackage{grffile} 

\usepackage{comment}

\usepackage{float}
\usepackage{dcolumn}
\usepackage{bm}
\usepackage[pdftex,colorlinks=true]{hyperref}
\hypersetup{
	colorlinks=true,
	linkcolor=blue,
	urlcolor=cyan,
}

\usepackage{color}
\definecolor{ForestGreen}{RGB}{34, 139, 34}

\newcommand{\ket}[1]{\left | #1 \right\rangle}

\begin{document}
		
	\title{Adaptive Trotterization for time-dependent Hamiltonian quantum dynamics \\
    using piecewise conservation laws}
	
	\author{Hongzheng Zhao}
	\email{hzhao@pku.edu.cn}
 \affiliation{School of Physics, Peking University, 100871 Beijing, China}
	\affiliation{Max Planck Institute for the Physics of Complex Systems, N\"{o}thnitzer Str.~38, 01187 Dresden, Germany}
 \affiliation{Institute for Quantum Optics and
Quantum Information, Technikerstraße 21a
6020 Innsbruck, Austria}
	
	\author{Marin Bukov}
	\affiliation{Max Planck Institute for the Physics of Complex Systems, N\"{o}thnitzer Str.~38, 01187 Dresden, Germany}

	\author{Markus Heyl}
	\affiliation{Max Planck Institute for the Physics of Complex Systems, N\"{o}thnitzer Str.~38, 01187 Dresden, Germany}
	\affiliation{Theoretical Physics III, Center for Electronic Correlations and Magnetism,
		Institute of Physics, University of Augsburg, 86135 Augsburg, Germany}
	
	\author{Roderich Moessner}
	\affiliation{Max Planck Institute for the Physics of Complex Systems, N\"{o}thnitzer Str.~38, 01187 Dresden, Germany}

	\date{\today}

\begin{abstract}
Digital quantum simulation relies on Trotterization to discretize time evolution into elementary quantum gates. On current quantum processors with notable gate imperfections, there is a critical tradeoff between improved  accuracy for finer timesteps, and increased error rate on account of the larger circuit depth. We present an adaptive Trotterization algorithm to cope with time-dependent Hamiltonians, where we propose 
a concept of {\it piecewise ``conserved" quantities} to estimate errors in 
the time evolution between two (nearby) points in time; these allow us to bound the  errors accumulated over the full simulation period.
They reduce to standard conservation laws in the case of time-independent Hamiltonians, for which we first developed an adaptive Trotterization scheme~[PRX Quantum 4, 030319]. We validate the algorithm for a time-dependent quantum spin chain, demonstrating that it can outperform the conventional Trotter algorithm with a fixed step size at a controlled error.
\end{abstract}
	\maketitle
\let\oldaddcontentsline\addcontentsline
\renewcommand{\addcontentsline}[3]{}

\textit{Introduction.---}
Simulating the time evolution of non-equilibrium quantum many-body systems poses a significant challenge for classical computers due to the exponentially large Hilbert space dimension~\cite{georgescu2014quantum,preskill2018quantum}. The rapid development of quantum processors holds the promise of resolving this key problem through digital quantum simulation (DQS)~\cite{blatt2012quantum,monroe2021programmable,dumitrescu2022dynamical,salathe2015digital,satzinger2021realizing,dborin2022simulating,moses2023race,chen2023continuous,kim2023evidence}.

In DQS, the continuous time evolution operator is discretized into elementary few-body quantum gates, a procedure known as Trotterization~\cite{suzuki1991general,berry2007efficient,poulin2014trotter,babbush2016exponentially,heyl2019quantum,tranter2019ordering,cirstoiu2020variational,bolens2021reinforcement,yao2021adaptive,lin2021real,richter2021simulating,mansuroglu2021variational,keever2022classically,pastori2022characterization,tepaske2022optimal,zhang2023low,mansuroglu2023problem,granet2023continuous}. However, due to their noncommutativity, Trotterization introduces errors, which can accumulate over longer simulation times. While a finer Trotter time step size $\delta t$ improves simulation precision, it also leads to increased circuit depth. In the current era of noisy intermediate-scale quantum (NISQ) processors, gate imperfections are inevitable, posing a significant challenge in improving the accuracy of DQS~\cite{preskill2018quantum}, especially in the absence of experimentally efficient error-correction schemes~\cite{postler2022demonstration,krinner2022realizing,chen2022calibrated}. Therefore, it is crucial to identify strategies for minimizing circuit depth while keeping the simulation error under control.

In previous work~\cite{zhao2023making}, we introduced a quantum algorithm, ADA-Trotter, allowing for adaptive step sizes $\delta t$ to optimize the usage of quantum gates for time-\textit{independent} Hamiltonians. 
By measuring the expectation values of energy and energy variance, $\delta t$ is maximized as long as errors in these conserved quantities remain bounded. According to the central limit theorem, ADA-Trotter ensures a correct energy distribution for generic non-integrable many-body systems. 
However, extending this formalism to time-dependent Hamiltonians $H(t)$ is a demanding challenge, since:
(i) Energy conservation is absent and hence it is a priori unclear how to define a criterion to adapt $\delta t$;
(ii) {without a static reference Hamiltonian the energy distribution is difficult to define, and hence the implication of the central limit theorem, mentioned above, is now elusive;
(iii) {generic many-body systems absorb energy from time-dependent modulations and may heat up (in the sense of the Eigenstate Thermalization Hypothesis~\cite{d2016quantum,abanin2019colloquium}) and eventually approach states with trivial correlations~\cite{lazarides2014equilibrium,abanin2015exponentially,kuwahara2016floquet}. It is not clear how to control the additional heating generated by the piecewise constant time-dependence of a  Trotterized  Hamiltonian compared to that of $H(t)$.}

In this work, we propose tADA-Trotter -- an adaptive algorithm for time-dependent quantum systems~\cite{poulin2011quantum,chen2021quantum,lau2021noisy,watkins2022time,ikeda2022minimum,an2022time,ikeda2023trotter24,kovalsky2023self}. To achieve this, we first discretize the time evolution into small time intervals $[t, t{+}\delta t]$. In each interval, the time evolution can be generated by an effective Hamiltonian $H_{[\infty]}^{t,\delta t}$.
Note that this Hamiltonian has only a parametric dependence on $t,\delta t$; it is time-independent once $t$ and $\delta t$ are fixed. Consequently, the expectation values of $H_{[\infty]}^{t,\delta t}$  and its higher moments coincide at the boundaries of this time window for perfect time evolution, a feature we will refer to as \textit{piecewise conservation laws}, cf.~Fig.~\ref{fig.schematic} (a). {In practice, $H_{[\infty]}^{t,\delta t}$ can be
approximated
using a perturbative Magnus expansion in the small time step $\delta t$~\cite{hochbruck2003magnus, alvermann2011high}. The Magnus and Trotterization approximations introduce errors in the piecewise conservation laws.} Our key finding is that, by constraining these errors, the step size can be adapted to reduce circuit depth while maintaining a given simulation accuracy for generic non-integrable many-body systems.

\begin{figure*}[t!]
 \includegraphics[width=\linewidth]{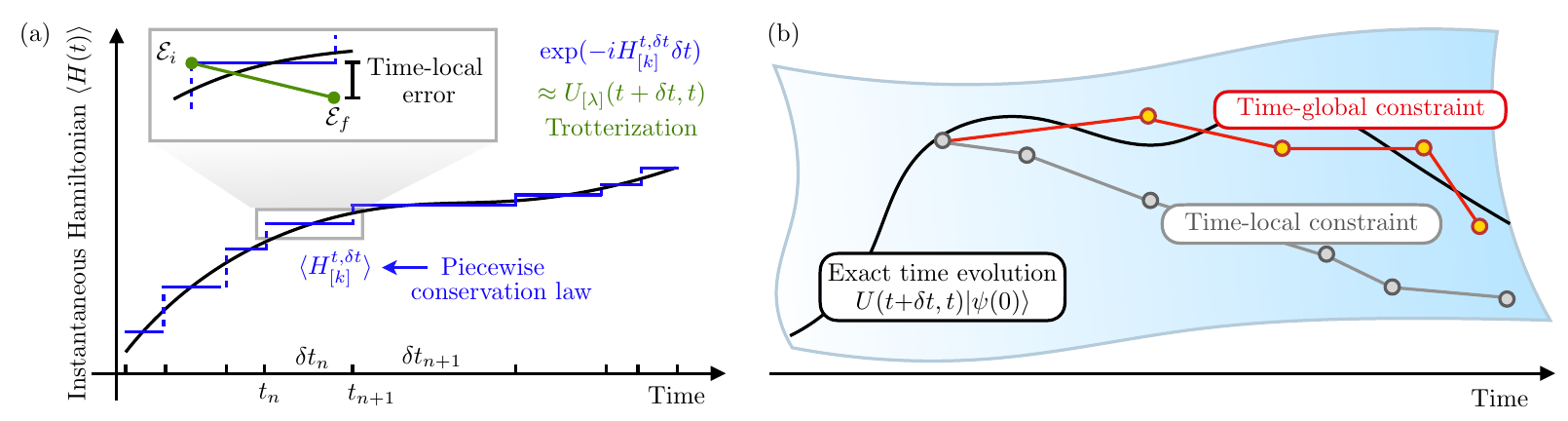}
		\caption{Schematics of tADA-Trotter for a time-dependent Hamiltonian. (a) {The expectation values of the piecewise conserved Hamiltonian $H_{[\infty]}^{t, \delta t}$ coincide at times $t$ and $t+\delta t$. {The exact expectation value of $H(t)$ is depicted as black line.} We use a Magnus expansion to approximate this conservation law  $H_{[\infty]}^{t,\delta t}{\approx }H_{[k]}^{t,\delta t}$(blue) and a Trotter decomposition (green) for the time evolution. Both approximations introduce errors to the time evolution.  We maximize $\delta t$ as long as errors in this conservation law
are bounded, i.e., deviations in the expectation value before ($\mathcal{E}_i$) and after ($\mathcal{E}_f$) the Trotterized evolution (green) should be small.} 
(b) {Schematic depiction of state space, showing the exact time
evolution (black line) as well as the approximate
state vector evolution calculated using tADA-Trotter under either
time-local (gray/black circles) or time-global (orange/red circles)
constraints}.{Trotter errors accumulate with the time-local constraint (grey), $\left|\mathcal{E}_f\left(t_n, \delta t_n\right)-\mathcal{E}_i\left(t_n, \delta t_n\right)\right|<d_{\mathcal{E}}$. This can be suppressed by using the time-global (accumulated) constraint (red), $\left|\sum_n\left[\mathcal{E}_f\left(t_n, \delta t_n\right)-\mathcal{E}_i\left(t_n, \delta t_n\right)\right]\right|<d'_{\mathcal{E}}$, leading to more reliable simulation with adaptive step sizes.}
}
\label{fig.schematic}
\end{figure*}

 {The adaptive step size is determined by measuring piecewise conservation laws using Trotterized time evolution.}
Subsequently, we introduce the concept of a \textit{time-global error}, which represents the accumulation of time-local errors in the piecewise conservation laws, cf. Fig.~\ref{fig.schematic}. To adapt the step size, we propose a feedback procedure that can bound both the local and global errors. 

Notably, Trotter-induced heating effects present in a local control scheme, where simulation errors significantly accumulate over time by making suboptimal choices of $\delta t$, may be efficiently suppressed by imposing constraints on the global errors, cf. Fig.~\ref{fig.schematic} (b). For time-independent systems, this globally constrained scheme reduces to the algorithm proposed in Ref.~\cite{zhao2023making}, enabling strict error bounds throughout the entire time evolution.
To determine the advantages of a globally constrained error, we perform numerical simulations of a quantum spin chain with a time-dependent  field. We also demonstrate that tADA-Trotter outperforms the conventional fixed-step Trotter, as depicted in Fig.~\ref{fig.main}. These findings highlight the superior potential of tADA-Trotter with a global constraint in minimizing circuit depth for DQS of time-dependent systems.

\textit{Piecewise conservation laws.---}The time evolution operator $U(t,t')$ follows the equation $\partial_t U(t,t'){=}-iH(t) U(t,t')$, where $H(t)$ represents the time-dependent Hamiltonian. Its solution is given by the time-ordered exponential $U(t+\delta t, t){=}\mathcal{T} \exp \left(-i\int_{t}^{t{+}\delta t} H(s) \mathrm ds\right)$, and the exact state evolves as $\ket{\phi(t{+}\delta t)}{=}U(t{+}\delta t,t)\ket{\phi(t)}$.

Formally rewriting $U(t{+}\delta t, t){=}\exp \left(-iH_{[\infty]}\delta t\right)$ indicates that the same time evolution can be generated by the static effective Hamiltonian $H_{[\infty]}$, where we drop its parametric dependence on $t,\delta t$ for simplicity. Hence, when $t$ and $\delta t$ are fixed, the expectation value of $H_{[\infty]}$, and its higher-order moments, coincide for the states $\ket{\phi(t{+}\delta t)}$ and $\ket{\phi(t)}$. We use these \textit{piecewise conservation laws} to adapt the Trotter step size $\delta t$.

The piecewise conserved Hamiltonian can be obtained through a Magnus expansion given by $H_{[\infty]}=i\delta t^{-1}\sum_{n=1}^{\infty} \Omega_n$, where the operator $\Omega_n\propto\delta t^n$. The explicit form of $H_{[\infty]}$ can be complicated as higher-order contributions typically involve nested commutators.
To eliminate the time-ordered integral in the Magnus expansion, we expand the time-dependence in Legendre polynomials, obtaining the concise expression for terms of lowest orders ~\cite{alvermann2011high}: $\Omega_{2m}{=}0$ for all even orders $2m$, and
\begin{eqnarray}
    \begin{aligned}
    \label{eq.approximation}
&\Omega_1{=}  A_1, \
\Omega_3{=} -\frac{1}{6}\left[A_1, A_2\right], 
 \Omega_5{=}  \frac{1}{60}\left[A_1,\left[A_1, A_3\right]\right]{-}\\
&\frac{1}{60}\left[A_2,\left[A_1, A_2\right]\right]
{+}\frac{1}{360}\left[A_1,\left[A_1,\left[A_1, A_2\right]\right]\right]{-}\frac{1}{30}\left[A_2, A_3\right],
\end{aligned}
\end{eqnarray}
where each operator $A_n^{t,\delta t}$ is defined as $A_n^{t,\delta t}=-i(2 n{-}1) \delta t \int_0^1 H(t{+}x \delta t) P_{n-1}(x) dx$. Here, $P_{n-1}$ denotes the shifted Legendre polynomials normalized to $(2n+1)\int_0^1 dsP_m(s)P_n(s){=}\delta_{mn}$, and $A_n^{t,\delta t}\propto\delta t^n$. For a sufficiently small time interval $\delta t$, this Magnus expansion can be truncated at a finite order $k$, resulting in an approximation of $H_{[\infty]}$ as $H_{[k]}{=}i\delta t^{-1}\sum_{n=1}^k\Omega_n$.

\textit{Trotterization.---}  
On real digital quantum devices, the exact time evolution operator $U(t{+}\delta t, t)$ for a smoothly varying Hamiltonian $H(t)$ is usually inaccessible. Thus, one needs to decompose the former into some elementary quantum gates using, e.g., Trotterization. For simplicity, we focus on the time-dependence {$H(t){=}g(t) G{+}f(t) F$ with smooth functions $g(t)$ and $f(t)$, and two generic non-commuting hermitian operators $F$ and $G$. Let us assume that quantum devices admit the exact implementation of unitaries of the form $\exp(-iC_1G)$ or $\exp(-iC_2F)$ where $C_{1,2}$ are arbitrary real numbers; while the implementation of linear combinations of $G$ and $F$, such as $\exp(-iC_1G{-}iC_2F)$}, is not feasible. 

We aim to approximate the target unitary operator up to a given order $\lambda$, such that
$
 U(t{+}\delta t,t) = U_{[\lambda]}(t{+}\delta t,t)
 {+}\mathcal{O}(\delta t^{\lambda}).$  A larger $\lambda$ leads to more accurate time evolution with smaller Trotter errors, but it also increases the circuit depth. {The number of gates generally scales exponentially in $\lambda$, but better decompositions with fewer number of exponentials may exist~\cite{ikeda2022minimum}}. In this work, we use $\lambda{=}3$ and the approximation can be obtained by the second-order Trotter formula, also known as the mid-point rule:
 \begin{equation}
 \begin{aligned}
 \label{eq.trotterU}
&U_{[3]}(t{+}\delta t,t){=}
\exp[{-ig(t{+}\delta t / 2) G\delta t/2}]\times\\
&\exp[{-if(t{+}\delta t / 2)F\delta t }] \exp[{-ig(t{+}\delta t / 2) G\delta t/2}].
\end{aligned}
\end{equation}

\textit{Adaptive algorithm.---}
The central concept of the adaptive algorithm is to maximize $\delta t$ while ensuring that the measurement outcome of the expectation value and variance of $H_{[\infty]}$ remain within pre-set tolerances. However, $H_{[\infty]}$ is generically increasingly non-local as contributions of increasing order $\Omega_n$ are involved, introducing a significant measurement overhead. Therefore, depending on the measurement accuracy and efficiency, one may need to truncate $H_{[\infty]}$ to a finite order $k$ to make the measurement procedure feasible on quantum computers~\cite{kokail2019self,mi2021information,naldesi2022fermionic}. Here, we consider a sufficiently large value for $k$, ensuring that errors in the piecewise conservation law are subdominant compared to the Trotter error. {In the limit $\delta t\to 0$, this can be satisfied as long as $k\geq\lambda$; we prove it using perturbation theory, see Sec.~\ref{sec.perturbative_error} for details. }

Below, we first introduce a time-local control scheme to adapt $\delta t$, which we find can involve severe heating effects. Then we propose a global control scheme to bound the accumulated errors and suppress heating.

In contrast to time-independent systems where conserved energies solely depend on the initial state $|\psi(0)\rangle$, the expectation values of {$H_{[k]}$} for generic time-dependent systems rely on the time-evolved state $|\psi(t)\rangle$. As a result, there are no universal (in time) reference expectation values known a priori for the piecewise conserved Hamiltonians. Nonetheless, we can leverage the capability of quantum processors to measure the expectation values, $\mathcal{E}_i (t,\delta t) {=} \langle \psi(t)| H_{[k]}|\psi(t)\rangle/L $ {with the system size $L$}, as a reasonable approximation to the conserved quantities for the true quantum state $|\phi(t)\rangle$. We maximize $\delta t$ such that the time-local error in the expectation value of $H_{[k]}$ remains below a threshold $d_\mathcal{E}$ (cf. Fig.~\ref{fig.schematic}(a)), i.e.,
\begin{equation}
|\mathcal{E}_f (t,\delta t)-\mathcal{E}_i (t,\delta t)|< d_{\mathcal{E}},
\end{equation}
where $\mathcal{E}_f (t,\delta t) {=} \langle \psi(t{+}\delta t)| {H_{[k]}}|\psi(t{+}\delta t)\rangle/L $ represents the expectation value of {$H_{[k]}$} after the Trotterized evolution, given by $|\psi(t{+}\delta t)\rangle {=} U_{[\lambda]}(t,\delta t) |\psi(t)\rangle$.
In the ideal case of $\lambda{\to}\infty$ and {$k{\to}\infty$}, we have $\mathcal{E}_f (t,\delta t){-}\mathcal{E}_i(t,\delta t){=}0$ by definition. However, for any finite value of $\lambda$ and {$k$}, this error does not vanish.

In addition, we also require the error in the variance to be bounded by the tolerance $ d_{\delta \mathcal{E}^2}$, i.e., $|\delta \mathcal{E}^2_f (t,\delta t){-}\delta \mathcal{E}^2_i (t,\delta t)|{<} d_{\delta \mathcal{E}^2}$, with 
\begin{eqnarray}
     \begin{aligned}
     &\delta \mathcal{E}^2_i(t,\delta t) {=} L^{-1} \langle \psi(t) | H_{[k]}^2(t,\delta t) | \psi(t)\rangle {-} L\mathcal{E}^2_i,\\
        &\delta \mathcal{E}^2_f(t,\delta t) {=} L^{-1} \langle \psi(t+\delta t) | H_{[k]}^2(t,\delta t) | \psi(t+\delta t)\rangle {-} L\mathcal{E}^2_f.
    \end{aligned}
\end{eqnarray}
According to the central limit theorem, constraining the errors in the lowest two moments of $H_{[k]}$ is sufficient to ensure the approximate conservation of its higher moments~\cite{hartmann2004gaussian,zhao2023making}. Therefore, the piecewise conservation of the Hamiltonian $H_{[k]}$ can be satisfied reasonably well, enabling reliable DQS of dynamics from time $t$ to $t{+}\delta t$.

Note that Trotter errors can accumulate in the time-evolved  $|\psi(t)\rangle$, leading to deviations of $\mathcal{E}_i$ and $\delta \mathcal{E}_i^2$, from the exact piecewise conservation laws. 
This effect is also present  for time-independent systems, where $H_{[k]}$ simplifies to a static Hamiltonian $H$. The energy constraint reduces to
$
| \langle \psi(t)| H|\psi(t)\rangle - \langle \psi(t{+}\delta t)|H|\psi(t{+}\delta t)\rangle|/L{<}d_{\mathcal{E}},
$
which accumulates and cannot be bounded for long simulation times. {Consequently, many-body systems tend to heat up and eventually become featureless. 
{This Trotter-induced heating can be {much more pronounced} in time-dependent systems, }leading to unstable DQS of time evolution over long periods.}

To address this challenge, we propose restrictions on the time-global errors, representing the accumulation of all time-local errors from previous steps:
\begin{equation}
\begin{aligned}
\label{eq.GlobalConstraints}
        &\left|\sum\nolimits_{n=1}^{m} [\mathcal{E}_f (t_n,\delta t_n)-\mathcal{E}_i (t_n,\delta t_n)]\right|< d'_\mathcal{E},\\
        &\left|\sum\nolimits_{n=1}^{m} [\delta \mathcal{E}^2_f (t_n,\delta t_n)-\delta \mathcal{E}^2_i (t_n,\delta t_n)]\right|< d'_{\delta \mathcal{E}^2}.
\end{aligned}
\end{equation}
{These conditions imply the time-local constraints, e.g., $|\mathcal{E}_f(t_m,\delta t_m) {-} \mathcal{E}_i(t_m,\delta t_m)| {<} 2d'_{\mathcal{E}}$~\footnote{To show this, we define $\Delta_{m-1}{=}\sum\nolimits_{n=1}^{m-1} \mathcal{E}_f (t_n,\delta t_n)-\mathcal{E}_i (t_n,\delta t_n),$ which satisfies $-d'_{\mathcal{E}}{<}\Delta_{m-1}{<}d'_{\mathcal{E}}$ with a positive $d'_{\mathcal{E}}$. For the next step, one has $-d'_{\mathcal{E}}<\Delta_{m-1}+\mathcal{E}_f (t_m,\delta t_m)-\mathcal{E}_i (t_m,\delta t_m)<d'_{\mathcal{E}}$, leading to $-2d'_{\mathcal{E}}<\mathcal{E}_f (t_m,\delta t_m)-\mathcal{E}_i (t_m,\delta t_m)<2d'_{\mathcal{E}}$}, but the converse is not true.  Therefore, information from the past time steps is used to select the current step size, such that the algorithm is capable of {automatically counteracting any accumulating Trotter-induced heating effects.} This global control is necessary to handle the lack of energy conservation in time-dependent systems.} 

We enforce these constraints via a feedback loop: initially, a large time step $\delta t_m$ is chosen. We then measure $\mathcal{E}_i$ and $\delta \mathcal{E}_i^2$ for the current quantum state $|\psi(t_m)\rangle$ and for the selected $\delta t_m$, as a prediction of the piecewise conserved quantities. We then implement the time evolution $U_{[\lambda]}(t_m,\delta t_m)$ on the quantum processor, yielding a candidate state $|\tilde \psi(t_m{+}\delta t_m)\rangle{=}U_{[\lambda]}(t_m,\delta t_m)|\psi(t_m)\rangle$. For this, we measure $\tilde{\mathcal{E}}_f$ and $\delta\tilde{\mathcal{E}}_f^2$. In case the measurement outcome violates the conditions of Eq.~\eqref{eq.GlobalConstraints}, a new smaller step size is proposed and the procedure restarts. 
 
{We use the bisection search method to find a new suitable $\delta t_m$. This can be efficiently implemented with a few trials whose number does not scale with system size and the truncation order $k$, cf. Ref.~\cite{zhao2023making} and Sec.~\ref{sec.costs}. }
{The extra measurement cost only depends polynomially on the system size and can be
further improved to logarithmic dependence by using classical shadows}~\cite{huang2020predicting,huang2021efficient,elben2023randomized}.
Once a suitable $\delta t_m$ has been found, we obtain the state $|\psi(t_m{+}\delta t_m)\rangle$ at the next time, and repeat the procedure.

 \begin{figure}
    \centering
\includegraphics[width=\linewidth]{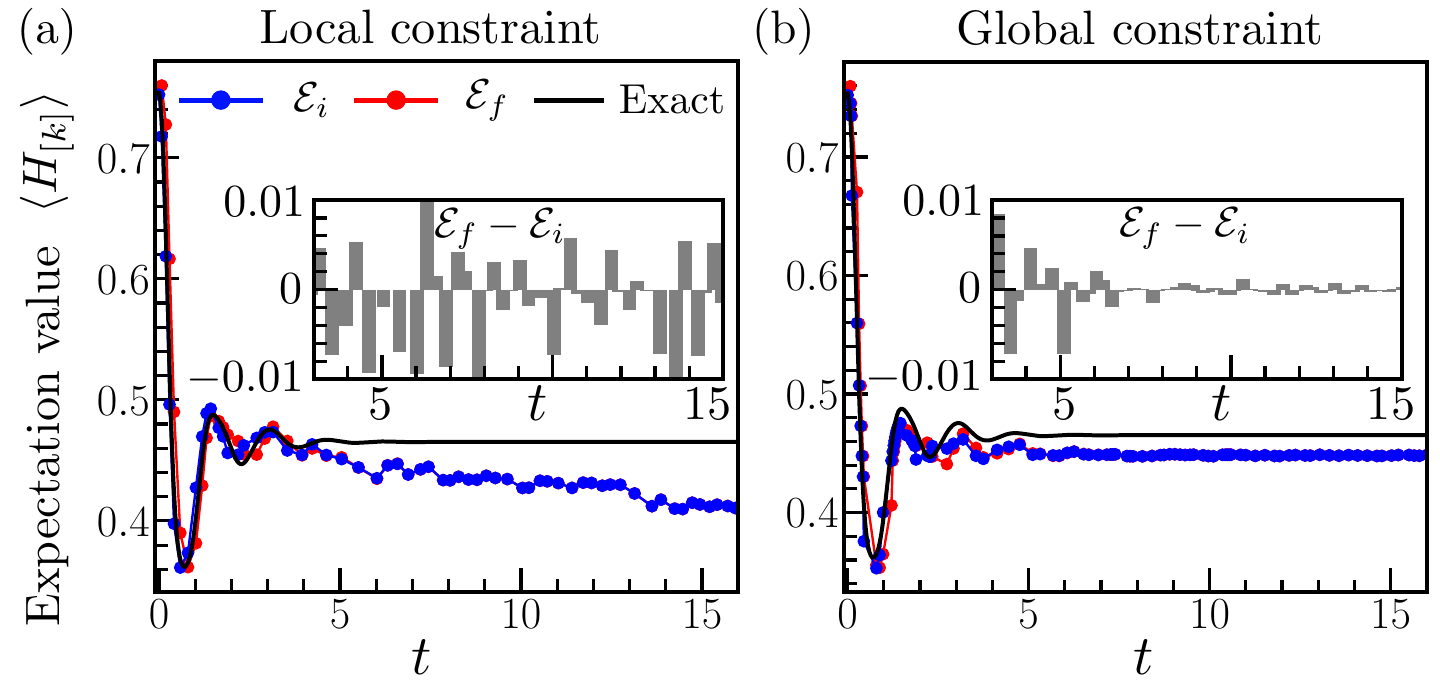}
    \caption{Comparison between the time-local and global constraint. (a) Local constraint with tolerances $d_{\mathcal{E}}=0.01$, $d_{\delta \mathcal{E}^2}=0.02,$ and $d'_{\mathcal{E}}=d'_{\delta \mathcal{E}^2}=\infty$. (b) Time-global errors are bounded, with the tolerance $d_{\mathcal{E}}=d_{\delta \mathcal{E}^2}=\infty,d'_{\mathcal{E}}=0.01, d'_{\delta \mathcal{E}^2}=0.02$. The following Hamiltonian parameters are used for numerical simulation, $J_z=1,h_x=1,h_z=0.5,\tau=1$,$\omega=4,L=16,\theta=2$. 
    }
    \label{fig.energy}
\end{figure}
\textit{Numerical simulation.---}
We next numerically compare  local and global constraint schemes and find  {better} performance of tADA-Trotter with  time-global control.

Although this algorithm is applicable to various models and initial states, for concreteness, we start from a product state $\ket{\psi(0)}{=}\exp(-i\theta\sum_j\sigma_j^x)\ket{\downarrow\dots\downarrow}$ and 
a non-integrable quantum Ising model, with Hamiltonian $H(t){=}g(t)H_x{+}f(t)H_z$ with
$
H_z{=} J_z \sum\nolimits_{j} \sigma_{j}^{z} \sigma_{j+1}^{z}+h_z \sum\nolimits_{j} \sigma_{j}^{z}, H_x{=}h_x \sum\nolimits_{j} \sigma_{j}^{x},$
where $\sigma_j^x$ and $\sigma_j^z$ are Pauli matrices acting on site $j$ of a chain consisting of $L$ sites with periodic boundary conditions. We consider a uniform coupling $J_z$, and transverse and longitudinal fields $h_x$ and $h_z$, respectively. 

We choose a static longitudinal field $f(t){=}1$ and an oscillating transverse field $g(t){=}\cos(\omega t)\exp(-t/\tau){+}1$ with nonzero mean,  frequency $\omega$, and exponentially decaying  amplitude. For  $t{\gg} \tau$, the system becomes effectively time-independent. Hence, this protocol contains different timescales and thus provides an ideal testbed for our algorithm. We employ Eq.~\eqref{eq.trotterU} to implement the Trotterized dynamics and truncate the piecewise conserved Hamiltonian to $H_{[k]}$ with $k=5$.

In Fig.~\ref{fig.energy}, we depict the expectation value of $H_{[k]}$ with the local and global control schemes. The exact solution  
varies at early times and becomes static at later times as expected. The predicted conserved value $\mathcal{E}_i$  at early times closely follows the exact solution. The expectation value $\mathcal{E}_f$ after implementing the Trotterized dynamics deviates from $\mathcal{E}_i$ very weakly in both cases.

A crucial difference occurs at later times. In Fig.~\ref{fig.energy} (a), for $t{>}5$ in units of the Ising coupling $J_z$, the predicted value $\mathcal{E}_i$ exhibits a noticeable drift towards zero, indicating the accumulation of Trotter errors. 
{Note, this Trotter-induced heating is not the same as the energy-non-conservation that goes along with $H(t)$~\cite{lazarides2014equilibrium,abanin2015exponentially,kuwahara2016floquet}. It happens because statistically, a stepsize increasing the system's entropy is more likely. To emphasize this point, we plot the time-local error $\mathcal{E}_f(t_n,\delta t_n) {-} \mathcal{E}_i(t_n,\delta t_n)$ at each time, and clearly, $\delta t$ is chosen in a way that negative values appear more frequently}. By contrast, when constraining the global errors according to Eq.~\eqref{eq.GlobalConstraints}, this deviation approximately centers around zero, indicating better controlled heating. Consequently, the overall drift  in $\mathcal{E}_i$ is suppressed. {
For a specific time $t_0$, as long as $t_0{\gg}\tau$ where the system becomes effectively time-independent, one can show that $\mathcal{E}_f(t_m,\delta t_m){=}\mathcal{E}_i(t_{m+1},\delta t_{m+1})$ for any $t_m{\ge} t_0$. Hence, Eq.~\eqref{eq.GlobalConstraints} reduces to
$\left| \mathcal{E}_i (t_m,\delta t_m){-}\mathcal{E}_i(t_{0},\delta t_{0})\right|{<} d'_\mathcal{E}$, strictly prohibiting the overall drift at long times as shown in Fig.~\ref{fig.energy} (b).}

\begin{figure}
    \centering
\includegraphics[width=\linewidth]{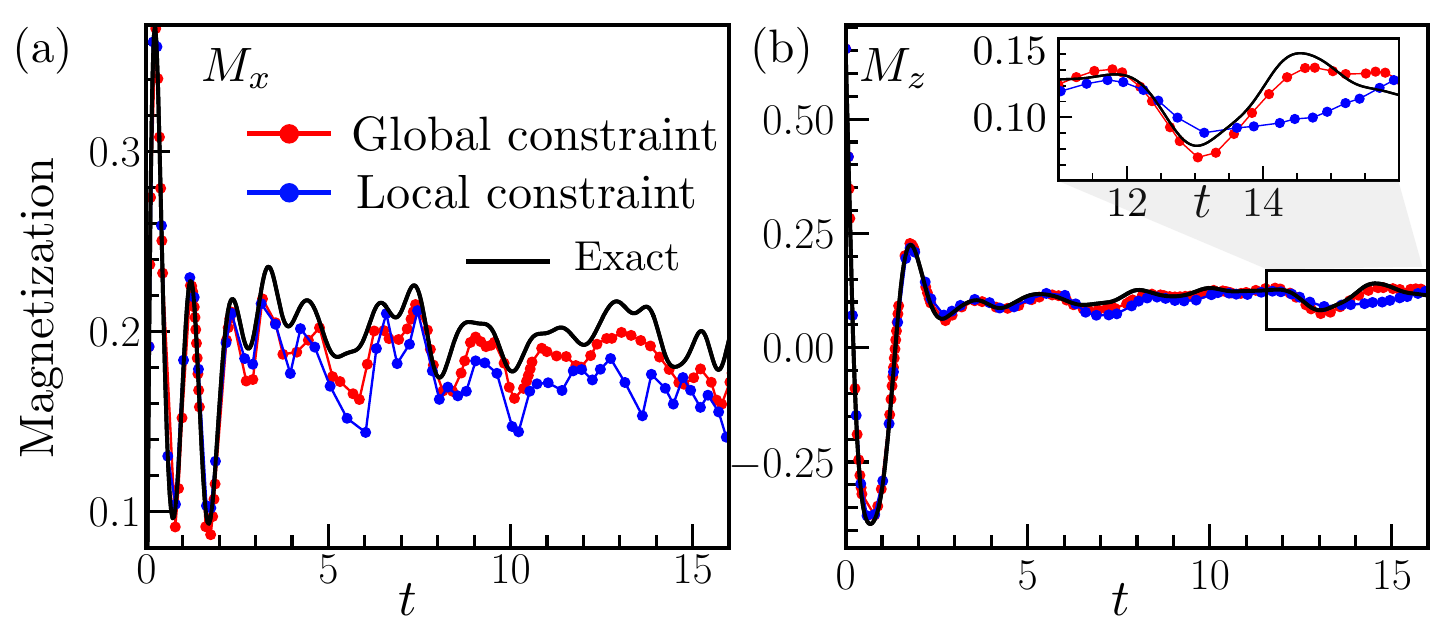}
    \caption{Global constraint leads to more stable simulation results than local constraint. Magnetization in $x$ (a) and $z$ (b) direction. We use the same parameters as in Fig.~\ref{fig.energy}. {The inset shows details of the dynamics in a small time window.}
    }
    \label{fig.observable}
\end{figure}
{Controlling errors in the expectation values of $H_{[k]}$} ensures accurate DQS of local observables. Fig.~\ref{fig.observable} shows the  magnetizations $M_\alpha=\sum_j\sigma_j^\alpha/L$ for $\alpha=x,z$,  demonstrating that  global constraint generally yields a more accurate simulation compared to the local constraint. 
In particular, in Fig.~\ref{fig.observable} (a) with the local constraint, significant errors arise for $t{>}5$. It corresponds to the time when notable deviation arises in the piecewise conserved quantities as shown in Fig.~\ref{fig.energy} (a). In contrast, the globally constrained data closely follows the exact solution for an extended period.

We now demonstrate that by constraining the time-global errors, tADA-Trotter achieves superior simulation precision compared to the fixed-step Trotter when the same total simulation time is reached. In the field $x(t)$ we select a  driving frequency $\omega{=} 0.8$ that is comparable to other local energy scales in the system. The characteristic decay timescale is chosen as $\tau{=}30$. With a total number of Trotter steps $N{=}100$, the achievable simulation time is approximately $t{\sim} 20$, during which the significant time-dependence in the Hamiltonian is still present.

\begin{figure}
	\includegraphics[width=0.85\linewidth]{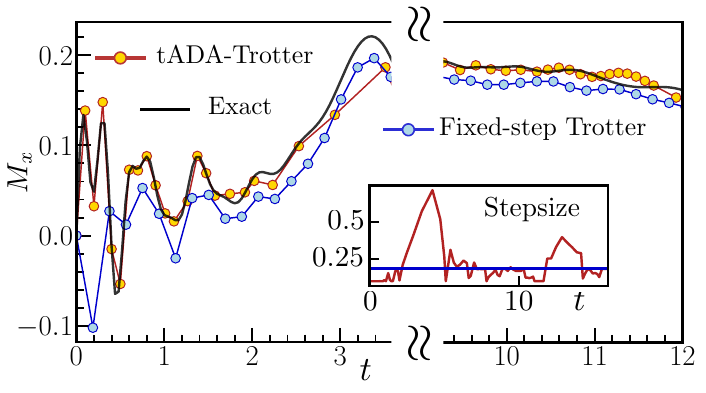}
	\caption{ Comparison between tADA-Trotter and fixed-step Trotter algorithms. Inset depicts the stepsize that varies in time. It takes larger values when external driving fields are weak. We use the following Hamiltonian parameters for numerical simulation, $J_z=1,h_x=3,h_z=0.5,\tau=30,\omega=0.8,L=18,\theta=2, d'_{\mathcal{E}}=0.03, d'_{\delta \mathcal{E}^2}=0.1$. 
	}
	\label{fig.main}
\end{figure}

In Fig.~\ref{fig.main}, $M_x{=}\sum_j\sigma_j^x/L$ is depicted with orange circles, which closely reproduces the exact solution (black). Simulation errors only become visible at later times, e.g., $t_m{>}11$. In contrast, the fixed-step Trotter ($\delta t_m{=}0.2$) already introduces substantial errors in the magnetization within a short time.
The Trotter step size (Fig.~\ref{fig.main} inset) fluctuates within approximately one order of magnitude $\delta t_m{\in} [0.1,0.7]$, highlighting the advantage and the flexibility of tADA-Trotter. Particularly, at early times, when the quantum state undergoes rapid changes under a strong driving field, smaller step sizes are employed ($\delta t_m{\approx} 0.1$). Conversely, when the driving $g(t)$ has relatively smaller values around $t_m{\approx} 4$ and $12$, the step size automatically increases to $\delta t{\approx} 0.7$ and $0.4$, respectively.

The time-global errors, Fig.~\ref{fig.errors},  remain bounded below the specified thresholds  for the majority of the time evolution. However, it should be noted that due to the tight tolerances at early times, the accumulated errors in the piecewise conserved quantities may occasionally exceed the bounds. This phenomenon can also cause the tADA-Trotter to ``freeze", {wherein it tends to select the smallest possible step size, which in this case is set to $0.1$.}

\textit{Discussion.---}
We have introduced the concept of 
piecewise conserved quantities and combined it with a time-global error constraint to devise an adaptive Trotterisation scheme to enable reliable DQS of generic time-dependent Hamiltonians. 
{In the presence of weak hardware noise, we expect that tADA-Trotter remains robust and can even self-correct the dissipation-induced errors~\footnote{We have explicitly demonstrated this for a time-independent Hamiltonian in Ref.~\cite{zhao2023making}}.} {We estimate that the shot overhead for a single step is around $10^4$, which can be efficiently implemented on current quantum processors with fast gate implementation, cf. details in Sec. \ref{sec.costs}.}

\begin{figure}
	\centering
	\includegraphics[width=0.85\linewidth]{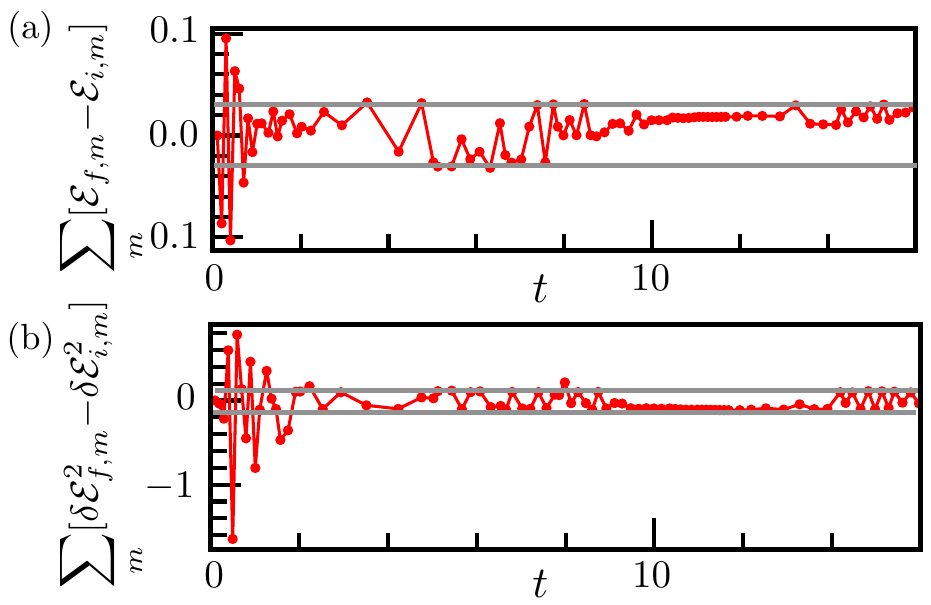}
	\caption{Time-global errors under the global constraint in the expectation value and variance of $H_{[k]}$. Errors are bounded for most of the time evolution. Same parameters as in Fig.~\ref{fig.main}. }
	\label{fig.errors}
\end{figure}

In a specific example, we demonstrate the superior performance of tADA-Trotter compared to the fixed-step Trotter method. The intricate interplay between external driving and quantum thermalization can result in highly complex many-body dynamics~\cite{kolodrubetz2017geometry,sugiura2021adiabatic,de2021colloquium,ho2023quantum}. Therefore, for future investigations, conducting a systematic benchmark of various algorithms across different models, initial states, and time-dependence would be of great value, as would be a comparison against higher-order truncation schemes.

Furthermore, the extension of piecewise conservation laws to open quantum systems to enable adaptive Trotter step sizes represents an intriguing open question~\cite{han2021experimental,de2021quantum,kamakari2022digital,vovk2022entanglement}. Additionally, considering the widespread use of Trotterization in classical numerical algorithms such as the time-evolving block decimation method, the application of tADA-Trotter to enhance the efficiency and accuracy of these methods holds significant potential.

\textit{Note added.---} During the completion of this work, we became aware of a relevant work exploring another adaptive algorithm for DQS of time evolution~\cite{ikeda2023trotter24}.

\textit{Acknowledgments}
We thank T. Ikeda for enlightening discussions. This work is in part supported by 
``The Fundamental Research Funds for the Central Universities, Peking University”, and by ``High-performance Computing Platform of Peking University” and the Deutsche Forschungsgemeinschaft  under  cluster of excellence ct.qmat (EXC 2147, project-id 390858490).
	This project has received funding from the European Research Council (ERC)
	under the European Unions Horizon 2020 research and innovation programme
	(grant agreement No. 853443).
MB was supported in part by the International Centre for Theoretical Sciences (ICTS) for participating in the program - Periodically and quasi-periodically driven complex systems (code: ICTS/pdcs2023/6).

\bibliography{Trotter}
 \let\addcontentsline\oldaddcontentsline
	\cleardoublepage
	\onecolumngrid
 \begin{center}
\textbf{\large{\textit{Supplementary Material} \\ \smallskip
	Adaptive Trotterization for time-dependent Hamiltonian quantum dynamics
using piecewise conservation laws}}\\
		\hfill \break
		\smallskip
	\end{center}
	
	\renewcommand{\thefigure}{S\arabic{figure}}
	\setcounter{figure}{0}
	\renewcommand{\theequation}{S.\arabic{equation}}
	\setcounter{equation}{0}
	\renewcommand{\thesection}{SM\;\arabic{section}}
	\setcounter{section}{0}
	\tableofcontents

\section{Perturbative analysis of errors}
\label{sec.perturbative_error}
In this section, by using a perturbative expansion, we show that errors in the piecewise conservation law are subdominant compared to the Trotter error if  $k\geq\lambda.$ 

At a given time $t$ and a Trotter step size $\delta t$, the exact
time evolution operator can be generated by the piecewise conserved quantity $H_{[\infty]}$,
 $U(t{+}\delta t, t){=}\exp \left(-iH_{[\infty]}\delta t\right)$. 
The target unitary operator can be Trotterized up to a given order $\lambda$, such that
$
 U(t{+}\delta t,t) = U_{[\lambda]}(t{+}\delta t,t)
 {+}\mathcal{O}(\delta t^{\lambda}).$ 
 One can formally rewrite the Trotterized unitary
  $U_{[\lambda]}(t{+}\delta t,t){=}\exp \left(-i\tilde{H}_{[\lambda]}\delta t\right)$ with 
  \begin{equation}
  \label{eq.lambda_expansion}
H_{[\infty]}=\tilde{H}_{[\lambda]}+\mathcal{O}(\delta t^{\lambda-1}),
  \end{equation}
  where we use $\sim$ to distinguish $\tilde{H}_{[\lambda]}$ with the approximated piecewise conserved quantity $H_{[k]}$, and we have
\begin{eqnarray}
      \label{eq.k_expansion}
      H_{[\infty]} = H_{[k]}+\mathcal{O}(\delta t^{k}).
\end{eqnarray} $H_{[k]}$ will be measured to adapt the step size. Suppose $\mathcal{E}_{i,k} = \langle \psi| {H_{[k]}}|\psi\rangle/L$ for a given state $\ket{\psi}$ 
and $\mathcal{E}_{f,k} = \langle \psi|U_{[\lambda]}^{\dagger} {H_{[k]}}U_{[\lambda]}|\psi\rangle/L $ represents the expectation value of ${H_{[k]}}$ after the Trotterized evolution. 
For $k\to\infty$, we have 
\begin{equation}
\begin{aligned}
\label{eq.e_expansion}
 \mathcal{E}_{f,\infty}=L^{-1} \langle \psi|U_{[\lambda]}^{\dagger} {H_{[\infty]}}U_{[\lambda]}|\psi\rangle  &= L^{-1}      \langle {\psi}|\exp[i\Tilde{H}_{[\lambda]}\delta t]H_{[\infty]}\exp[-i\Tilde{H}_{[\lambda]}\delta t]|{\psi}\rangle\\
         &= L^{-1} \langle {\psi}|H_{[\infty]}+i\delta t[\Tilde{H}_{[\lambda]},H_{[\infty]}]+\frac{1}{2}(i\delta t)^2[\Tilde{H}_{[\lambda]},[\Tilde{H}_{[\lambda]},H_{[\infty]}]]|{{\psi}}\rangle +\dots\\
        &=\mathcal{E}_{i,\infty}+L^{-1}  \langle {\psi}|i\delta t[\Tilde{H}_{[\lambda]},H_{[\infty]}]+\frac{1}{2}(i\delta t)^2[\Tilde{H}_{[\lambda]},[\Tilde{H}_{[\lambda]},H_{[\infty]}]]|{{\psi}}\rangle +\dots
\end{aligned}
\end{equation}
where we use the formula 
\begin{eqnarray}
\label{eq.BCH}
    e^A B e^{-A}=B+[A, B]+\frac{1}{2}[A,[A, B]]+\ldots,
\end{eqnarray} to perform the perturbative expansion in $\delta t$
and neglect contributions from higher orders.
By using Eq.~\eqref{eq.lambda_expansion}, one has $[\Tilde{H}_{[\lambda]},H_{[\infty]}]\sim\delta t^{\lambda-1}$, and $[\Tilde{H}_{[\lambda]},[\Tilde{H}_{[\lambda]},H_{[\infty]}]]\sim\delta t^{\lambda-1}$. Therefore the leading order contribution to the change in the piecewise conservation law reads
\begin{eqnarray}
\label{eq.E_error_infty}
   \mathcal{E}_{f,\infty}-\mathcal{E}_{i,\infty}=L^{-1} \langle {\psi}|i\delta t[\Tilde{H}_{[\lambda]},H_{[\infty]}]|{{\psi}}\rangle+\mathcal{O}(\delta t^{\lambda+1})\sim \delta t^{\lambda}.
\end{eqnarray}
Ideally, if one can measure the expectation value of $H_{[\infty]}$, $ \mathcal{E}_{f,\infty}-\mathcal{E}_{i,\infty}$ will be used to adapt the step size according to tADA-Trotter algorithm. However, for generic many-body systems, $H_{\infty}$ is a highly non-local and complicated operator. We now truncate it to a finite order $k$ such that $H_{[k]}$ can have a local structure and can be measured in practice. However, such truncation introduces errors in addition to the Trotter error (its dependence on $\delta t$ is purely determined by $\lambda$).
Now we define this additional error for a finite $k$ as $\Delta=(\mathcal{E}_{f,\infty}-\mathcal{E}_{i,\infty})-(\mathcal{E}_{f,k}-\mathcal{E}_{i,k})$, and we aim to find the condition such that $\Delta$ is subdominant than $\mathcal{E}_{f,\infty}-\mathcal{E}_{i,\infty}$. Such a comparison is generally a challenging task since these quantities highly depend on the state and the time-dependent Hamiltonian. Instead, we analyse the scaling of $\Delta$ versus $\delta t$, and compare its asymptotic behavior between $\Delta$ and $\mathcal{E}_{f,\infty}-\mathcal{E}_{i,\infty}$ for $\delta t\to 0$. By definition,
\begin{eqnarray}
\begin{aligned}
\label{eq.definition_Delta}
    \Delta &=   (\mathcal{E}_{f,\infty}-\mathcal{E}_{i,\infty})-(\mathcal{E}_{f,k}-\mathcal{E}_{i,k})= 
   \langle \psi|U_{[\lambda]}^{\dagger} (H_{[\infty]}-H_{[k]})U_{[\lambda]}|\psi\rangle-\langle \psi|(H_{[\infty]}-H_{[k]})|\psi\rangle.
   \end{aligned}
\end{eqnarray}
  By using Eq.~\eqref{eq.BCH}, we expand it perturbatively in $\delta t$,
  \begin{eqnarray}
\begin{aligned}
  \Delta= \langle {\psi}|i\delta t[\Tilde{H}_{[\lambda]},(H_{[\infty]}-H_{[k]})]+\frac{1}{2}(i\delta t)^2[\Tilde{H}_{[\lambda]},[\Tilde{H}_{[\lambda]},(H_{[\infty]}-H_{[k]})]]|{{\psi}}\rangle +\dots,
\end{aligned}
\end{eqnarray} where the leading order contribution scales as $\delta t^{k+1}$ according to Eq.~\eqref{eq.k_expansion}. Hence, to ensure that that errors in the expectation value of piecewise conserved quantity are
subdominant compared to the Trotter error, one needs $\mathcal{O}(\delta t^{k+1})<\mathcal{O}(\delta t^{\lambda})$, or equivalently, $k\geq\lambda.$ 

Now we argue the same condition suffices to ensure the variance of piecewise conserved quantity is dominated by Trotter error. The variance is defined as
\begin{eqnarray}
     \begin{aligned}
     \delta \mathcal{E}^2_{i,k} {=} L^{-1} \langle \psi | H_{[k]}^2| \psi\rangle {-} L\mathcal{E}^2_{i,k}, \delta \mathcal{E}^2_{f,k} {=} L^{-1} \langle \psi |U_{[\lambda]}^{\dagger} H_{[k]}^2 U_{[\lambda]}| \psi\rangle {-} L\mathcal{E}^2_{f,k}.
    \end{aligned}
\end{eqnarray}
For $k\to\infty$, one has
\begin{eqnarray}
\begin{aligned}
     \delta \mathcal{E}^2_{f,\infty}-\delta \mathcal{E}^2_{i,\infty} &= L^{-1}\left(\langle \psi |U_{[\lambda]}^{\dagger} H_{[\infty]}^2 U_{[\lambda]}| \psi\rangle-\langle \psi | H_{[\infty]}^2 | \psi\rangle\right)-L\left(\mathcal{E}^2_{f,\infty}- \mathcal{E}^2_{i,\infty}\right)\\
    &\approx L^{-1}\langle {\psi}|i\delta t[\Tilde{H}_{[\lambda]},H^2_{[\infty]}]|{{\psi}}\rangle-L\left(\mathcal{E}_{f,\infty}- \mathcal{E}_{i,\infty}\right)\left(\mathcal{E}_{f,\infty}+ \mathcal{E}_{i,\infty}\right).
\end{aligned}
\end{eqnarray}
By using Eq.~\eqref{eq.lambda_expansion}, Eq.~\eqref{eq.E_error_infty} and the fact $\left(\mathcal{E}_{f,\infty}+ \mathcal{E}_{i,\infty}\right)\sim \delta t^0$, one obtains $\delta \mathcal{E}^2_{f,\infty}-\delta \mathcal{E}^2_{i,\infty} \sim \delta t^{\lambda}$, which exhibits the same scaling as Eq.~\eqref{eq.E_error_infty}. Once we truncate $k$ to a finite number, one obtains the additional error for the variance
\begin{eqnarray}
\begin{aligned}
\label{eq.var_error}
          &(\delta \mathcal{E}^2_{f,\infty}-\delta \mathcal{E}^2_{i,\infty}) -  (\delta \mathcal{E}^2_{f,k}-\delta \mathcal{E}^2_{i,k} )\\&=L^{-1}\left(\langle {\psi}|U_{[\lambda]}^{\dagger}(H^2_{[\infty]}-H^2_{[k]})U_{[\lambda]}|{{\psi}}\rangle-\langle {\psi}|(H^2_{[\infty]}-H^2_{[k]})|{{\psi}}\rangle\right)-L\left[\left(\mathcal{E}^2_{f,\infty}- \mathcal{E}^2_{i,\infty}\right)-\left(\mathcal{E}^2_{f,k}- \mathcal{E}^2_{i,k}\right)\right]
\end{aligned}
\end{eqnarray}
With Eq.~\eqref{eq.k_expansion} and Eq.~\eqref{eq.BCH} we know the leading order contribution in the first term scales as $\delta t^{k+1}$. To obtain the scaling for the second term, we first use Eq.~\eqref{eq.e_expansion} for a finite $k$ and approximate $\mathcal{E}_{f,k}$ as
$
\mathcal{E}_{f,k}\approx\mathcal{E}_{i,k}+ \langle {\psi}|i\delta t[\Tilde{H}_{[\lambda]},H_{[k]}]\ket{\psi}/L,$ hence we have 
$
\mathcal{E}^2_{f,k}\approx\mathcal{E}^2_{i,k}+ 2i\delta t\mathcal{E}_{i,k}\langle {\psi}|[\Tilde{H}_{[\lambda]},H_{[k]}]\ket{\psi}/L,$
Therefore the second term in Eq.~\eqref{eq.var_error} becomes
\begin{eqnarray}
\begin{aligned}
\left(\mathcal{E}^2_{f,\infty}- \mathcal{E}^2_{i,\infty}\right)-\left(\mathcal{E}^2_{f,k}- \mathcal{E}^2_{i,k}\right)&={2i\delta t}{L^{-1}} \left(\mathcal{E}_{i,\infty}\langle\psi|[\tilde{H}_{\lambda},H_{[\infty]}]|\psi\rangle-\mathcal{E}_{i,k}\langle\psi|[\tilde{H}_{\lambda},H_{[k]}]|\psi\rangle\right)\\
&={2i\delta t}{L^{-1}} \left(\mathcal{E}_{i,k}\langle\psi|[\tilde{H}_{\lambda},H_{[\infty]}-H_{[k]}]|\psi\rangle+(\mathcal{E}_{i,\infty}-\mathcal{E}_{i,k})\langle\psi|[\tilde{H}_{\lambda},H_{[\infty]}]|\psi\rangle\right).
\end{aligned}
\end{eqnarray}
By noticing $\mathcal{E}_{i,k}$ scales as $\delta t^0$, $\mathcal{E}_{i,\infty}-\mathcal{E}_{i,k}\sim\delta t^k$ and also using Eq.~\eqref{eq.k_expansion} and Eq.~\eqref{eq.lambda_expansion}, one can show that the leading order contribution to the equation above also scales as $\delta t^{k+1}$. Consequently, to ensure that that errors in the expectation value and variance of the piecewise conservation law are
subdominant compared to the Trotter error asymptotically for a small $\delta t$, one needs  $k\geq\lambda.$ 

It is worth noting that the argument above requires a small $\delta t$ for Trotterized dynamics, implying that the circuit depth to reach certain target time can be large and unrealistic in the NISQ era. For any practical purpose, the optimal $k$ and $\lambda$ may highly depend on the target time-dependent systems, the measurement resources and gate fidelity in specific experimental setups. 

\section{Numerical simulation with different $k$}
\label{sec.differentK}
Sec.~\ref{sec.perturbative_error} suggests that $k\geq\lambda$ would be a suitable choice to ensure Trotter error is the only dominant error. For the Trotter decomposition with $\lambda=3$ as considered in the main text, the minimal $k$ to satisfy this condition would be $k=3$. In the main text, we fix $k=5$ to ensure that $\Delta$ defined in Eq.~\eqref{eq.definition_Delta} can be negligible. However, according to Eq.~\eqref{eq.approximation}, a large $k$ introduces extra measurement overheads. In this section we numerically show that a smaller $k$ can also generate sufficiently accurate Trotterized time-evolution with the global constraints, therefore the measurement overheads can be significantly reduced.

\begin{figure}[h]
\includegraphics[width=0.8\linewidth]{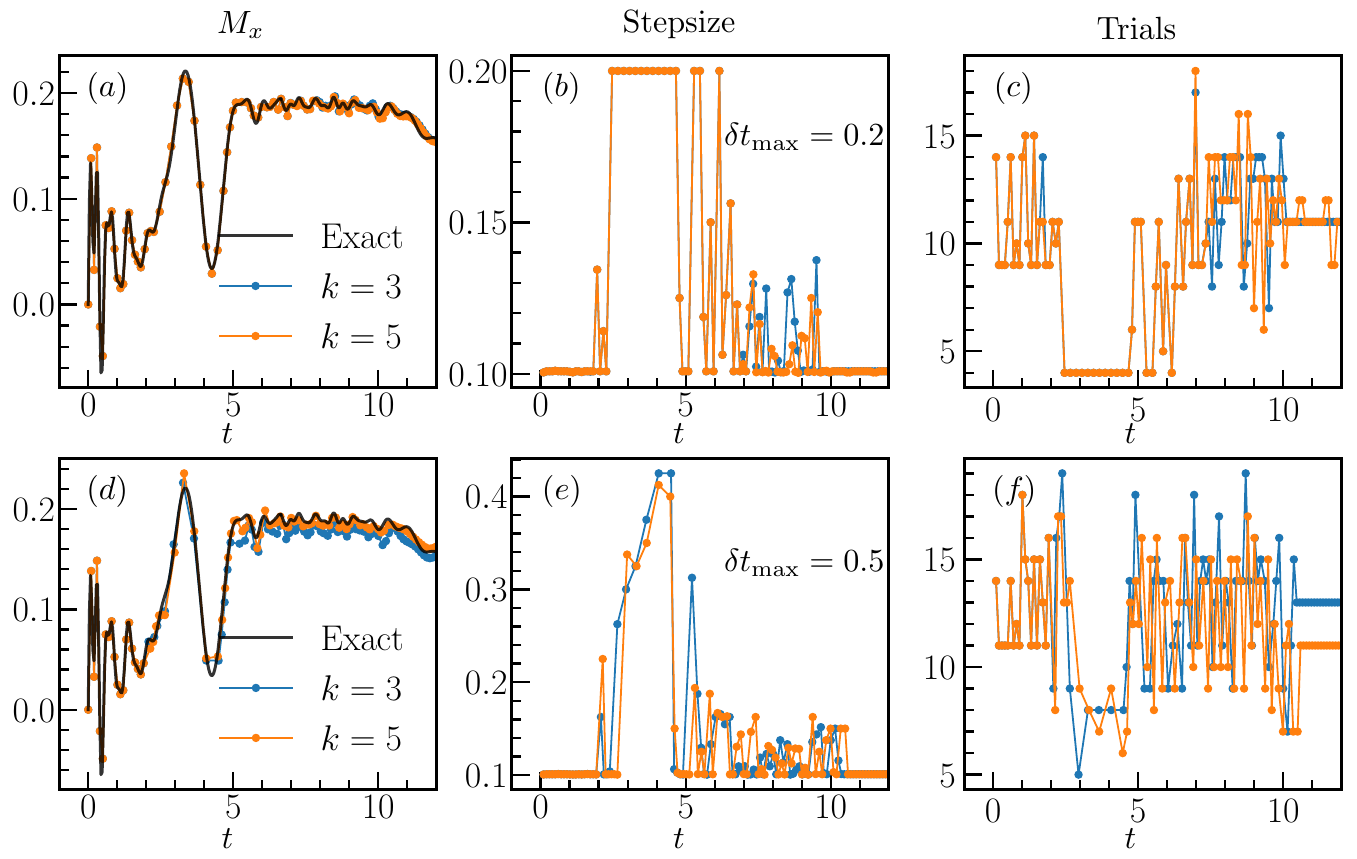}
		\caption{ Comparison between tADA-Trotter with different truncation order $k$ for the piecewise conservation law. (a) and (d) depict the dynamics of the magnetization: Both $k=3$ and $5$ lead to accurate simulation results compared with the exact solution(black). Panels (b) and (e) show the step sizes selected from tADA-Trotter: (b) uses a smaller maximum step size ($\delta t_{\text{max}}=0.2$) when the bisection search is implemented. Hence, errors in the piecewise conservation law $\Delta$ is subdominant and $k=3$ and $k=5$ lead to almost the same simulation outcomes. A larger $\delta t_{\text{max}}$ is used for (e), and differences in the selection of the step sizes are clearly visible for $k=3$ and $5$. (c) and (f) depict the number of trials that have been implemented for the bisection search. On average this algorithm requires 10 trials before determining the optimal step size. 
  We use the following Hamiltonian parameters for numerical simulation, $J_z=1,h_x=3,h_z=0.5,\tau=30,\omega=0.8,L=16,\theta=2, d'_{\mathcal{E}}=0.01, d'_{\delta \mathcal{E}^2}=0.02$. 
}
		\label{fig.compare_k35}
	\end{figure}
In Fig.~\ref{fig.compare_k35}, we first compare the simulation results with $k=3$ (blue) and $5$ (orange). For panels (a) (b) (c) we use $\delta t_{\text{max}}=0.2$ as the largest allowed step size for tADA-Trotter, whereas for (d)(e)(f) we use $\delta t_{\text{max}}=0.5$. Since the overall step sizes shown in panel (b) are relatively small, at each step, $H_{[k]}$ is almost the same for $k=3$ and $k=5$. Therefore the decision of $\delta t$ in panel (b) and the resulting dynamics in the magnetization are also almost the same, especially at early times. For (d)(e)(f), because larger step sizes are allowed, the difference in $H_{[k]}$ becomes visible and step sizes may be chosen differently for $k=3$ and $5$. Both simulation results follow closely with the exact solution, but the measurement overhead can be substantially smaller for $k=3$.

 \begin{figure}[h] \includegraphics[width=0.8\linewidth]{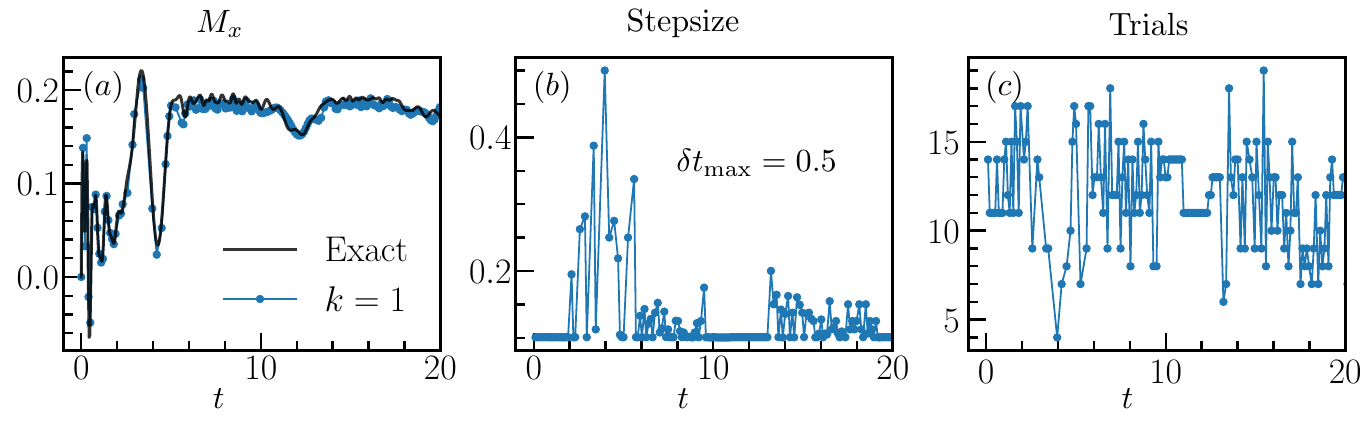}
		\caption{$k=1$ can also generate accurate Trotterized dynamics although the interplay between the error in the approximated piecewise conservation law $\Delta$ and the Trotter error becomes difficult to control. (a): dynamics of the magnetization.  (b): step sizes selected from tADA-Trotter. (c): the number of trials that have been implemented for the bisection search. On average this algorithm requires 10 trials before determining the optimal step size. 
  We use the following Hamiltonian parameters for numerical simulation, $J_z=1,h_x=3,h_z=0.5,\tau=30,\omega=0.8,L=16,\theta=2, d'_{\mathcal{E}}=0.01, d'_{\delta \mathcal{E}^2}=0.02$. 
}\label{fig.compare_k1}
	\end{figure}

The asymptotic limit $\delta t\to 0$ implies a divergent circuit depth to reach certain physical simulation time $t$. Hence, in practice, one normally uses larger step sizes on NISQ devices. In this case, even if the condition $k\geq \lambda$ is not satisfied, tADA-Trotter may still be able to generate reliable simulation of many-body dynamics. We demonstrate it in Fig.~\ref{fig.compare_k1} where $k=1$ is used. However, this observation is not universal and the performance of tADA-Trotter can highly rely on specific time-dependence of the system.

\section{Costs}
\label{sec.costs}
\subsection{Measurement of conservation laws}
tADA-Trotter introduces a measurement overhead to obtain expectation values of the piecewise conserved quantities. However, for a time-dependent and local Hamiltonian $H(t)$, $H_{[k]}$ with a finite $k$ will always remain local. Hence, measuring the expectation value of $H_{[k]}$ involves terms linear in systems size $L$ and measuring $H_{[k]}^2$ only involves terms quadratic in $L$. Indeed, a small value of $k$ can already generate reliable tADA-Trotter simulation results as shown in Sec.~\ref{sec.differentK}. Furthermore, the system size dependence can be improved based on randomized measurements and classical shadows as recently introduced ~\cite{huang2020predicting,huang2021efficient,elben2023randomized}. For instance, the measurement cost for the $H_{[k]}$ can be $L$-independent and the cost for the Hamiltonian variance has been improved to depend only logarithmically on the system size~\cite{huang2020predicting}. The measurement shot noise normally scales as $1/\sqrt{N}$ where $N$ denotes the number of measurement shots. Therefore, to reach certain accuracy $d'_{\delta \mathcal{E}}$ , $N$ roughly scales as $1/{d'}^2_{\delta \mathcal{E}}$. In principle, one can always perform more measurement shots to increase this accuracy. However, in practice, depending on the specific experimental setup, one would estimate the total measurement budgets and roughly distribute them over each Trotter step to preset $d'_{\delta \mathcal{E}}$.

{In terms of actual numbers, to reach the simulation accuracy of Fig.~\ref{fig.main}, we estimate that the shot overhead would be $10^3$ in a state-of-the-art quantum simulators for a single trial. This estimate is based on recent work which implemented a variational quantum simulation to identify the ground state of a target Hamiltonian on a trapped ion quantum simulator~\cite{kokail2019self}. There, to quantify the variational state, measurement of energy and energy variance were both performed. There, $10^2$ shots are needed to obtain a good estimate of the energy for a given state, which is very efficient. Measurement of energy variance is more costly: To ensure that the variational state is sufficiently close to the ground state, the authors performed around $4*10^4$ shots to resolve the energy gap (which is around 2 in units of their local energy scale) between the ground state and first excited state. $4*10^4$ shots are shown to be sufficient to determine an energy variance around 1.2 for a given variational state. Energy variance density is around $10^{-1}$, which is around the same accuracy required to obtain the results in our Fig.~\ref{fig.main}. In Ref.~\cite{huang2020predicting}, the authors further reduce this shot number to $2*10^3$ by using derandomized classical shadow methods. }

{The Hamiltonians studied in those works are 2-local, i.e., they contain single-site and two-site spin-1/2 operators. In our algorithm, if we choose the order $k=1$ or 3 to approximate conservation laws, the effective Hamiltonian is also 2-local. Hence, the shot overhead for energy and energy variance is roughly the same for a given trial state in our algorithm. With the bisection search method, the number of trials for a single Trotter step is 5-10 (see discussion in the next section), hence, the shot overhead for a single step is around $10^4$.}

{On more established quantum processors with efficient gate implementation, for instance superconducting qubits, the extra runtime for these overheads is truly negligible. One layer of CNOT gates roughly costs $10^{-7}$ seconds on IBM quantum processors, and a circuit involving hundreds of layers of CNOT for Tottered dynamics only costs $10^{-5}$ seconds. Even with $10^4$ shot overheads, a single step of tADA-Trotter only takes less than 1 second. }

{Therefore, our algorithm is definitely within the current reach of quantum processors and it provides the great potential to substantially improve the accuracy of Tottered dynamics. }

\subsection{Number of trials}
To determine the optimal step size, we currently use a bisection search algorithm, originally proposed in our previous work~\cite{zhao2023making}. {To reach certain search accuracy $\epsilon$, it introduces extra $\log (\epsilon)$ overheads. For tolerances considered in this work, tADA-Trotter} uses around 10 trials on average before identifying the optimal step size, see Fig.~\ref{fig.compare_k35} and Fig.~\ref{fig.compare_k1}. This trail number does not increase for larger system sizes and a larger truncation order $k$ (cf. panels (c) and (f) in Fig.~\ref{fig.compare_k35}), permitting the implementation of tADA-Trotter on real quantum devices that may have access to a much larger number of qubits. 

We are currently developing a technique dubbed ``Energy-Landscape Estimation" to further reduce the number of trials. By exploiting the continuity and analyticity of the Magnus expansion of the conserved quantity as a function of Trotter step size, the dependence of the energy and its variance can be efficiently estimated by only using five or even smaller trials. This method can be applied to both time-dependent and -independent systems and will further improve the efficiency of the tADA-Trotter.

\subsection{Simulation time versus Trotter step}
{Whenever a destructive measurement is performed, one needs to restart the experiment from the beginning and run the circuit again. Hence, the total simulation time on actual devices scales as $n^2$ where $n$ is the number of Trotter steps. Note, this is indeed the same scaling as the fixed step Trotter algorithm, because a measurement is performed after each Trotter step to obtain the transient dynamics during the time-evolution. Therefore, this does not introduce extra overhead.}
 
\end{document}